\documentclass[12pt,preprint2]{aastex}

\slugcomment{Draft}

\shorttitle{The Race to Build SMBHs}
\shortauthors{Tyler, Janus, and Santos-Noble}

\begin{document}

\title{The Race to Build Supermassive Black Holes}

\author{C. Tyler, B. Janus, D. Santos-Noble}
\affil{Department of Physics, Fort Lewis College,
    Durango, CO 81301}
\email{tyler\_c@fortlewis.edu}

\begin{abstract}
The high redshifts of the most distant known quasars, and
the best estimates of their black hole masses, require that
supermassive black holes (SMBHs) must have formed very early in
history.  Several mechanisms for creating and growing these holes
have been proposed.  Here we present an evaluation of the
timescales needed for various critical processes in order to
discriminate between the proposed scenarios.  We find in particular
that mergers alone are not able to grow the black holes at a
sufficient rate.  Accretion models offer a solution
and we use accretion timescales to constrain the manner
in which the black hole was first formed.  This analysis
implies, but does not require, the action of some
unconventional process.
\end{abstract}


\section{Introduction}  \label{introduction}

Central supermassive black holes are a common feature to
galaxies today, but which came first, the black hole or
the galaxy?  Conventional thinking would suggest that the the
first generation of stars evolved into black holes, which
have subsequently settled to the centers of
their host galaxies, merged, and accreted gas.
But this idea, in which central black holes form inside
pre-existing galaxies, has recently earned some scrutiny.
First, the discovery of increasingly high
redshift quasars requires a surprisingly early formation of
the black holes (see, for example, \citet{SDSS1} and
\citet{SDSS2}).  Second, a large quasar sample shows no evidence of
black holes growing in mass with decreasing redshift \citep{vest02,diet02}.
So we are left to consider the possibility that either the
central black holes formed before their host galaxies,
or they grew to maturity very quickly within them.  Either way,
they have grown little since the quasar epoch.

The most distant known quasar lies at $z = 6.41$,
with a central black hole of mass
$M_{\bullet} = 3 \times 10^9~{\rm M}_{\sun}$
\citep{will03}.  In the $\Lambda$CDM cosmology observed
by WMAP \citep{WMAP}, with $\Omega_{\Lambda} = 0.73$,
$\Omega_{\rm m} = 0.27$, and $H_{\rm 0} = 71~{\rm 
km~s^{-1}~Mpc^{-1}}$, this redshift corresponds to a time when the
universe was only $880~{\rm Myr}$ old.  For the present work,
we will take this as the time to beat:  $3$ billion solar masses
in $880$ million years.

In the past year, two separate HST studies have cited
evidence for intermediate mass black holes (IMBHs) in the
centers of globular clusters:  a $4000~{\rm M}_{\sun}$ hole
in M15 \citep{vdm02}, and a $20\,000~{\rm M}_{\sun}$ hole in
Andromeda's G1 cluster \citep{gebh02}.  This is the lastest
and strongest evidence for IMBHs, but there is additional
evidence, and good theoretical motivation as well; see
\citet{vdm03} for a comprehensive review.  IMBHs are
widely believed to be a necessary growth stage for SMBHs.
In section \ref{flowchart} of this paper, we will review the
major proposed routes to the formation of a SMBH, all of which
include an IMBH phase, from which time the question
is simply one of growth.

We start in Section \ref{flowchart} with a flowchart of
avenues leading to the creation of a SMBH.  In Section
\ref{timescales}, we examine the timescales for each
needed process.  We conclude in Section \ref{conclusions}
by discussing how realistic each avenue is in light of these
timescales.

\section{Flowchart}  \label{flowchart}

\begin{figure*}
\includegraphics[width=7in,height=7in]{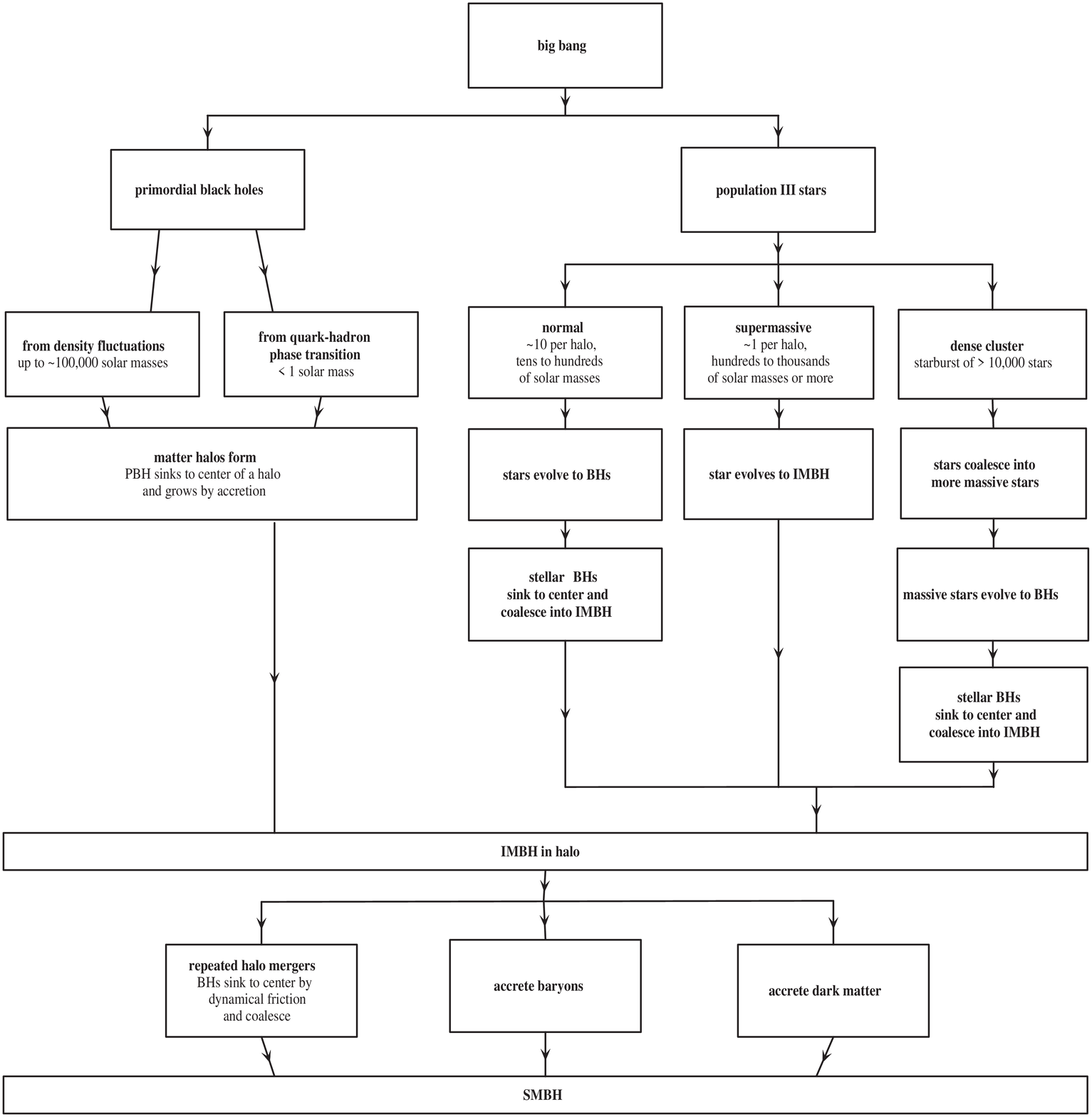}
\caption{The various ways to build the earliest
supermassive black holes in the universe.
\label{flow}}
\end{figure*}

There are essentially four proposed families of models leading
to the formation of IMBHs, and two or three ways to grow them.
These approaches are depicted in figure \ref{flow} and
discussed in turn below.

1.  The black holes may be \emph{primordial}, in which case
they formed from primordial density variations before big bang
nucleosynthesis.

Primordial black holes (PBHs) have been studied extensively,
and the most direct mechanism for their creation is the collapse
of gaussian random density fluctuations \citep{carr74}.
These holes come from horizon scale (or smaller) modes,
and therefore their masses are determined by their time
of formation.  In the radiation dominated early universe,
\begin{equation}
M_{\bullet} \simeq 10^5
\left( \frac{t}{\rm s} \right)~{\rm M}_{\sun} \,\,. 
\label{pbh_time}
\end{equation}
But in order to preserve the successful BBN prediction of
light element abundances, there must be no significant rate
of PBH formation once nucleosynthesis begins, and therefore the
PBHs are capped at intermediate mass.  In addition, \citet{carr94}
have pointed out that, given a small scalar spectral index
$-$ $n_s = 0.93 \pm 0.03$ was recently observed in the CMB
\citep{WMAP} $-$ PBHs from density inhomogeneities should
only have formed in quantities too small to be of interest.

A more promising, and perhaps inevitable mechanism for forming
PBHs also exists, in which the collapse is triggered by 
``color holes'' at the quark-hadron phase transition \citep{craw82}.
However, because this occurred at $\sim 10^{-6}~{\rm s}$, these PBHs
would be smaller than $\sim 1~{\rm M}_{\sun}$ by eq.
\ref{pbh_time}, and would remain as collisionless dark
matter today, rather than collecting into larger black holes.
(Interestingly, \citet{hawk96} shows evidence for such
PBHs in the microlensing of distant quasars, in numbers
comparable to that needed to account for dark matter.)

2.  Normal \emph{population III stars} formed at
$\sim 100~{\rm M}_{\sun}$, evolved to black holes,
and merged at the center of their small dark matter halos.

This is perhaps the least exotic way to create IMBHs, and
at this point there is very strong support for the process
in the form of numerical simulations of structure formation
\citep{abel00,yosh03}.  These simulations include the relevant
atomic and molecular processes in the first gas clouds,
particularly cooling by rotation of molecular hydrogen,
superimposed on CDM halo evolution.  They find that
$\lesssim 10$ stars of $\sim 100~{\rm M}_{\sun}$ form in halos
of $\sim 10^6~{\rm M}_{\sun}$, engaging $\lesssim 1\%$
of the system's baryonic matter.

Because the cooling of population III stars is hindered
by the lack of metals, these first stars would be more
massive than those allowed by fragmented star formation today.
\citet{hege02} have shown that these massive stars will evolve
into black holes containing a significant fraction of the star's
initial mass (this fraction varies with the stellar mass, and is
of order $50\%$), unless the stars are in the mass range
$140~{\rm M}_{\sun} < M < 260~{\rm M}_{\sun}$, in which
case they are completely disrupted when they go supernova.
Given the small initial size of such a cosmologically young
halo ($\sim 10~\rm{pc}$), the holes sink to the center and
presumably merge into a single IMBH there.

3.  \emph{Supermassive stars} may have been the first baryonic
objects to form.  If so, they would have evolved rapidly
into IMBHs.

Supermassive stars (SMSs), as a class of objects, span from 
$10^3$ to $10^8~{\rm M}_{\sun}$, although the first generation
of them would reside at the lower end of that range
(\citet{st} provides a comprehensive introduction
on the topic).  A $10^3~{\rm M}_{\sun}$ SMS has a lifetime of
$\lesssim 10^6$ years (SMS lifetimes range down to 10 years),
at the end of which, it undergoes relativistic collapse to
a black hole in a matter of seconds \citep{new01}.  This
collapse sends $\sim 90\%$ of the star's original mass
into the black hole, with the remaining $10\%$ forming an
accretion disk \citep{shap02,shib02}.

The radius of a $10^3~{\rm M}_{\sun}$ SMS,
$R_{\rm SMS} \approx 0.1~{\rm pc}$,
is around $10\%$ of the virial radius for the earliest halos
containing gas cool enough to collapse.  It is therefore
reasonable to suspect that these stars should appear in 
early halos, especially since population III stars
produce ultraviolet radation which splits hydrogen molecules,
thus destroying the only coolant which could collapse smaller
stars.  The collapsing gas might produce enough radiation pressure
to prevent fragmentation \citep{baum99}.  Another possibility
is that the first stars formed dense clusters which merged
into one or more SMSs.  However, whether or not supermassive
stars ended the cosmological dark ages remains a matter
of speculation.

4.  If \emph{dense clusters} of stars emerged from early
star forming regions, then the cluster stars may have merged
with each other to create massive stars (say, several hundred
solar masses), which then evolved to massive seed black holes.
This idea has been mapped out recently by \citet{ebis01},
and we will not repeat it here.  However, for the sake
of timescales, which are the focus of this paper, this
scenario is nearly the same as the population III
star scenario discussed above, because the merging occurs
during the lifetime of the cluster's massive stars.

Each of the paths in figure \ref{flow} sees a phase with
an IMBH imbedded in a halo, at which point accretion and
halo mergers take over to achieve heavier black holes.

The merger scenario merges central black holes hierarchically
along with the host halos, growing both together.  It is
clear that merging IMBHs and SMBHs are real, interesting
phenomena deserving study.  For example, \citet{milo01}
have simulated the whole process, and shown that coalescing
SMBHs remove the steep cusps that CDM simulations generate,
thus matching observed galaxy rotation curves.  But it remained
to be shown here whether or not mergers are prevalent enough to
sufficiently increase black hole mass.

Gas accretion is another major way to grow IMBHs into SMBHs.
The belief that quasars are powered by accretion at the Eddington
luminosity gives this idea credibility.  But over what fraction of
cosmic history did these holes accrete?  Some authors suppose
that Eddington limited accretion is continuous since the black
hole first formed \citep{haim01}, and thus  its mass grows rapidly.
Others have proposed scenarios in which accretion is
negligible over the history of the BH \citep{mada01}, such that
$150~{\rm M}_{\sun}$ holes that formed in the first episode
of star formation still populate our Galactic bulge today.

Another possibility is that the BH feeds on dark matter, in which
case the halo and the IMBH evolve together.  \citet{macm02}
proposed a model in which the halo and hole grow self-similarly,
with both analytical and numerical results \citep{macm03}
supporting their idea.  This model is particularly interesting
because it also simultaneously explains both the spectral
index for the galaxy power spectrum, $n \approx -2$,
and the $M_{\bullet} \propto \sigma_{\rm bulge}^4$
relationship observed to be universal in galaxy centers
\citep{ferr00,gebh00,trem02}.

All of these methods certainly contribute to the growth of
massive black holes.  In the next section we will
use timescales to help constrain which methods could
realistically dominate the process of forming SMBHs in
less than a gigayear.

\section{Timescales}  \label{timescales}

First, we examine the evolution to the IMBH phase, as depicted
in Figure \ref{flow}.  If the BH is primordial, then it can
be up to $10^5~{\rm M}_{\sun}$ in the very early universe.
Its growth by merger or accretion then awaits the formation of
matter clumps to surround the PBH.  These could begin to arrive
as early as photon decoupling, so timescales alone will not
strongly constrain this possibility unless the PBH in question
started at $< 1~\rm{M}_{\sun}$, as it would if it formed during
the quark-hadron phase transition.  In this case the BH seed
has a very long way to grow, which is barely possible by
accretion, as will be shown in subsection \ref{accretion}.

The simulations of population III star formation of
\citet{yosh03} produce some very useful information.
The earliest simulated star forming region occurs at
$z = 32$, with stars forming at $10^{-3}~{\rm M_{\sun}\,
yr^{-1}}$ per comoving ${\rm Mpc}^3$, increasing to
$0.1~{\rm M_{\sun}\, yr^{-1}\, Mpc^{-3}}$ at $z = 7$,
and holding constant thereafter \citep{hern03}.

Molecular hydrogen coolant forms in these clouds in
$\sim 30~{\rm Myrs}$.  By $z = 23$, the gas pressure has
diminished and gas clumps begin to contract.  Halos
are $\sim 5 \times 10^5$ to $2 \times 10^6~{\rm M}_{\sun}$
by this time, and the largest virialized star forming
clouds are $10^5~{\rm M}_{\sun}$.  The gas takes an
approximately isothermal density profile, with constant
density inside a core radius of $\sim 10~{\rm pc}$.

The virial temperature is a few thousand Kelvins.
A simple application of the virial theorem gives a virial
radius of $45~{\rm pc}$, and the corresponding circular orbital
speed is $10~{\rm km~s^{-1}}$.

This, then, is the starting point for population III stars
leading to IMBHs, with star formation beginning in earnest at
$z = 23$.  To convert redshifts to times in a $\Lambda$CDM
cosmology with $\Omega_{\rm 0} = 1$ \citep{peacock},
\begin{equation}
t_{\rm age} = \frac{2}{3} H_{\rm 0}^{-1}
\frac{{\rm sinh}^{-1}
% \left( \sqrt{ \frac{|\Omega_{\rm m} - 1|}
% {(1 + z)^3 \Omega_{\rm m}}} \right) }
\sqrt{ |\Omega_{\rm m} - 1| 
(1 + z)^{-3} \Omega_{\rm m}^{-1} } }
{\sqrt{|\Omega_{\rm m} - 1|}}
\label{z-t1}
\end{equation}
and with WMAP values inserted, this is
\begin{equation}
t_{\rm age} = 10.8 \, {\rm sinh}^{-1}
\left( \sqrt{2.7 \, (1 + z)^{-3}} \right) \, {\rm Gyr} \,.
\label{z-t2}
\end{equation}
So gas collapses into the first stars at
$t_{\rm age} \approx 0.15~{\rm Gyr}$.

These stars collapse in the free fall time,
\begin{equation}
t_{\rm ff} = \sqrt{ \frac{3 \pi}{32 \, G \rho} } \,, 
\label{tff}
\end{equation}
which is about $30~{\rm Myr}$ at the densities
observed in simulations by \citet{abel00}.  They
evolve to black holes (or disruptive supernovae)
in a few million years, leaving $\sim 0.7~{\rm Gyr}$
remaining to grow into a quasar power source.

Using the dynamical friction formula from \citet{chan43},
quoted here from \citet{bt}, we can find the time needed
for these population III remnant BHs to sink to the center
of their host halo.  Treating the BHs as point particles,
\begin{equation}
t_{\rm df} = \frac{1.17}{G \, {\rm ln} \, \Lambda} \,
\frac{v r_{\rm 0}^2}{M_{\bullet}} \, ,
\label{chandra_df}
\end{equation}
where $\rm{ln} \, \Lambda$ is called the Coulomb
logarithm, which characterizes the distance range
perpendicular to the path of motion over which the
frictional encounters occur (for a recent study of
this parameter, see \citet{spin03}).  Using the
values mentioned above for the parameters in this
equation, and choosing $\Lambda \approx 20$ (this choice is
common and doesn't affect the outcome much because of the
logarithm), the dynamical friction timescale relevant here is
at least $\sim 0.1~{\rm Gyr}$, bringing the total down to
$\sim 0.6~{\rm Gyr}$ available for BH growth.  The coalescence
time of these holes is not well known, and could be long.

If the star formation yields one or more supermassive stars,
rather than multiple normal population III stars,
then the evolution to a BH is faster than for normal stars
\citep{st}.  The dynamical friction time to merge smaller
stellar BHs is also avoided with SMSs, so in the SMS case,
we have $\sim 0.7~{\rm Gyr}$ remaining to grow a SMBH.

In the case of an early-forming dense star cluster in
which runaway collisions merge the stars, the timing remains
essentially the same.  One waits until the cluster stars
form, merge, and evolve to one or more BHs.
The time available to become supermassive
is still at most $0.7~\rm{Gyr}$.

\subsection{Dynamical Friction} \label{df}

We consider the case of a satellite subhalo halo falling into
a larger halo.  Each halo is modeled as a singular isothermal
sphere, so that $\rho \propto r^{-2}$ and orbital
velocities within a halo are constant.  We seek a formulation
that tracks the friction until the satellite's central
BH reaches the new center.

Smaller halos formed earlier in the CDM hierarchy, when the
universe was denser, and therefore they are more tightly
bound.  So as a subhalo sinks, we assume it holds itself
together, until it gets close to the center of
the larger halo.  There, which halo an orbiting particle
belongs to becomes ambiguous.  This occurs roughly when
a particle's distance from its original host subhalo
is on order of the separation between the two halo centers.
This separation distance is $r$, a radial coordinate
extending from the center of the larger halo.

The internal speed within an isothermal halo is
\begin{equation}
v = \sqrt{\frac{GM}{R}} \, .
\label{v_sis}
\end{equation}
The square root in this equation is helpful because it
reduces the dependence that velocities have on a halo
size $R$, which is generally not well known.

The mass which can be associated with the sinking
satellite is then
\begin{equation}
M_{\rm{s}} \simeq M_r + M_{\bullet} = 
\frac{v_{\rm{s}}^2 r}{G} + M_{\bullet} \, ,
\label{m_subhalo}
\end{equation}
where $M_r$ is the mass of the satellite halo residing
inside a distance equal to $r$ around it, $M_{\bullet}$ is
the satellite subhalo's central BH mass, and $v_{\rm{s}}$
is the orbital velocity inside the satellite.  Choosing to use
$r$ in this way assumes that the isothermal satellite halo
extends to distances larger than some initial infall radius:
$R_{\rm{s}} \geq r_i$.

The angular momentum of the satellite in the larger
halo is $\vec{L} = \vec{r} \times M_{\rm{s}} \vec{v}$,
where $\vec{v}$ is the orbital velocity within the larger halo.
Dynamical friction induces a torque
$\vec{\tau} = \vec{r} \times \vec{F} =
\vec{r} \times M_{\rm{s}} (d{\vec{v}}/dt)$.

Following the standard derivation presented in
\citet{bt}, the statement of $\vec{\tau} = d{\vec{L}}/dt$
leads to the equation
\begin{equation}
r \frac{dr}{dt} = - 0.428 \,
\frac{G M_{\rm{s}}}{v} \, \rm{ln} \, \Lambda \, .
\label{bt_df}
\end{equation}
We next define 3 constants as follows:
$C_1 = (0.428/v) \, \rm{ln} \, \Lambda$,
$C_2 = C_1 G M_{\bullet}$, and $C_3 = -C_1 v_{\rm{s}}^2$.
Inserting the satellite mass from eq. \ref{m_subhalo},
we have
\begin{equation}
\frac{dr}{dt} + C_2 r^{-1} = C_3 \, .
\label{diffeq}
\end{equation}

\begin{figure}[t]
\plotone{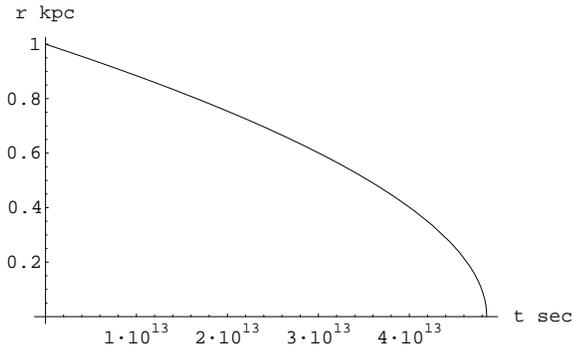}
\caption{Dynamical friction infall profile as calculated in
section \ref{df}.  In this figure we have exaggerated the
curve at the end for clarity by using an unrealistically
large BH mass in the satellite subhalo.
\label{roft}}
\end{figure}

Eq. \ref{diffeq} is a first-order, linear, ordinary
differential equation, but the closed-form solution is
prohibitively long and complicated due to the $r^{-1}$ factor
in the second term.  A numerical solution readily
parameterizes the sinking orbit $r(t)$, and a root can be
found for the time $t_{\rm{df}}$ at which $r \rightarrow 0$.
One needs only to specify a starting radius for the
satellite halo, $r_i \equiv r(0)$.  An example orbit
decay curve computed in this way is given in figure
\ref{roft}.

This dynamical friction model approximately doubles the
time spent spiralling inward, relative to the Chandrasekhar
formula for an infalling point particle, eq.
\ref{chandra_df}.  \citet{colp99} found the same
result using both $N$-body simulations and the theory of
linear response.  A subsequent paper by \citet{taff03}
gives an analytical approximation for NFW halos
\citep{nfw}, also finding that the decay time
is increased by a factor of a few if the satellite has less
than $\sim 10\%$ of the larger halo's mass.  In the mass
range of interest for the present work, all these models
essentially agree; however, since we wish to follow
length and mass scales ranging down to a central BH,
we have chosen to use the method described above.

\subsection{Mergers} \label{mergers}

We first consider the viability of black hole merging as
a means to grow SMBHs.  As halos merge over cosmic history,
their central BHs sink to the new center by dynamical
friction, and subsequently coalesce into a larger BH.

The usual way to track mergers is by $N$-body simulation.  We
begin with such a simulation, and calculate the total mass of
subhalos that can merge into one large halo between the
epoch of the first massive BHs and the early quasar epoch.  We
subject this calculation to the constraint that the merging 
subhalos must sink to the center in the time allowed
between the two epochs.

This can become a very complicated calculation, because
the halo population is continuously changing over a long
period of cosmic time.  To make the problem tractable,
we make a series of major approximations.  We describe
them along the way, and discuss them afterward.  We
demonstrate first that with our assumptions, mergers
are \emph{not} sufficient to grow SMBHs, regardless of
how the BHs originally formed.  Second, we show that a
more accurate calculation to replace our simplifying
assumptions would only strengthen that conclusion.

The argument begins with the numerical simulations of
\citet{volo03}.  They follow a $4 \times 10^{13}~\rm{M}_{\sun}$
halo today back in time to all its subhalos, presenting
halo mass spectra at multiple redshifts.  For our purposes,
the results they present at $z = 5$ and $z = 20$ are most important,
approximating the time of the first quasars ($z = 6.41$) and the
first IMBHs in halos (after $z = 23$).  We'll address the impact
of these approximations in redshift and total mass
after the calculation.

The halo functions are nearly power law in the mass range
of interest.  For $M$ in $\rm{M}_{\sun}$ units, at $z = 5$
\begin{equation}
\frac{dN}{dM} = 10^{14} \, M^{- \frac{7}{3} } \, ,
\label{dNdM5}
\end{equation}
and at $z = 20$
\begin{equation}
\frac{dN}{dM} = 1.6 \times 10^{16} \, M^{-2.9} \, .
\label{dNdM20}
\end{equation}

Consider the largest halo in existence as the period of interest
begins.  Imagine (conservatively) that all other subhalos are
available to merge into this one, and (also conservatively) that
all of these subhalos contain central BHs of $0.1\%$ of the parent
galaxy's luminous mass, as observed in large galaxies today
\citep{ferr02}.  According to WMAP, the ratio of baryonic matter
to total matter is $0.17$ \citep{WMAP}, so conservatively we may say
that a central BH has $0.02\%$ of the host subhalo's mass.  Then
every subhalo massive enough to dynamically sink to the center
in the available time does sink, leading to a more massive
combined halo.

If all the IMBHs from all the combined subhalos coalesce quickly
(again, highly conservative), then the new SMBH mass is
$0.02\%$ of the total mass of all subhalos with mass
$M \gtrsim M_{\rm{s,min}}$, the minimum satellite mass needed
for dynamical friction to sink a subhalo by the early quasar era.

However, during the merging process, the largest halo
is gaining mass and the number of massive halos is
increasing.  To evade this problem, we suppose
that the halo mass spectrum at the \emph{end} of the
merging process (at $z = 5$) applies throughout the
merging process.  Our calculations show that in most
cases, this assumtion conservatively exaggerates
the final SMBH mass by more than an order of magnitude.

Using eq. \ref{dNdM5}, the total mass of the
quasar's SMBH is
\begin{equation}
M_{\bullet} = 0.0002 \int_{M_{\rm{s,min}}}^{\infty}
10^{14} \, M^{-\frac{4}{3}} \, dM \, .
\label{int_dNdM}
\end{equation}
The lower limit, $M_{\rm{s,min}}$, is determined by
dynamical friction as described in section \ref{df}.  We
follow the infalling halo mass, and transition to the (smaller)
infalling BH mass as the two holes close in, achieving results 
closely consistent with other models that include mass loss
but not a central BH \citep{colp99}.

In the dynamical friction calculation, three parameters must
be chosen.  We take the internal velocity of the larger
halo to be $v \approx 100~\rm{km~s^{-1}}$, noting that for a fixed
dynamical sinking time (fixed by the time available for BH
growth), the final SMBH mass obeys $M_{\bullet} \propto
v^{-1/3}$.  So choosing a larger velocity would  only very
slightly reduce the final SMBH mass by reducing the
number of subhalos which sink in the allowed time.

The other needed quantities are radii:  $r_i$, the initial radius
from which the satellite halo begins its descent, and $R_{\rm{s}}$,
the size of the satellite subhalo.  The former is roughly
the impact parameter of the colliding subhalos, and the
latter indicates the subhalo's concentration factor.  Both
radii will clearly vary from one encounter to the next, so we
choose several representative values.  The results demonstrate
that any reasonable (in fact, conservative) choice
of these numbers leads to the same conclusion, that
merging black holes take too long.

As shown earlier, the timescale between the first
IMBH and a quasar's SMBH depends on how the BH formed.
A primordial BH has the age of the universe at the time of
the most distant quasar's emission known today, or
$0.88~\rm{Gyr}$.  Population III stars and
supermassive stars have $0.7~\rm{Gyr}$.  If the
population III stars merged within their original halo first,
then they have $0.6~\rm{Gyr}$ to grow.

We fix the sinking time at the appropriate value
and compute the total BH mass that can be assembled
by halo mergers, neglecting the time needed for coalescence.
Our results are listed in Table \ref{tab_merger}, which
lists a maximum SMBH mass for various choices of
$R_{\rm{s}}$ and $r_i$.  No choice brings $M_{\bullet}$
up near $3 \times 10^9~\rm{M}_{\sun}$, the value needed
to explain the most distant known quasars.

\begin{table}[t]
\begin{center}
\begin{tabular}{c c c c} \hline
{$t$ (Gyr)} & {$R_{\rm{s}}$ (kpc)} & {$r_i$ (kpc)}
& {$M_{\bullet}$ ($\rm{M}_{\sun}$)} \\
\hline \hline
$0.88$ & $10$ & $10$ & $4.8 \times 10^7$ \\
& $10$ & $1$ & $1.0 \times 10^8$ \\
(PBH) & $10$ & $0.1$ & $2.2 \times 10^8$ \\
& $1$ & $1$ & $2.2 \times 10^8$ \\
& $1$ & $0.1$ & $4.8 \times 10^8$ \\
\hline
$0.7$ & $10$ & $10$ & $4.4 \times 10^7$ \\
& $10$ & $1$ & $9.5 \times 10^7$ \\
(pop III, & $10$ & $0.1$ & $2.1 \times 10^8$ \\
cluster, & $1$ & $1$ & $2.1 \times 10^8$ \\
or SMS)& $1$ & $0.1$ & $4.4 \times 10^8$ \\
\hline
$0.6$ & $10$ & $10$ & $4.2 \times 10^7$ \\
& $10$ & $1$ & $9.0 \times 10^7$ \\
(merged & $10$ & $0.1$ & $2.0 \times 10^8$ \\
pop III) & $1$ & $1$ & $1.9 \times 10^8$ \\
& $1$ & $0.1$ & $4.2 \times 10^8$ \\
\hline
\end{tabular}
\end{center}
\caption{SMBH masses achieved through halo mergers by the
early quasar epoch.  $R_{\rm{s}}$ is a characteristic size
for the satellite subhalos, and $r_i$ is the radius from
which they are assumed to begin their infall.
$t$ is the time allowed for growth.}
\label{tab_merger}
\end{table}

One is led to conclude that mergers alone are not sufficient
to grow the largest SMBHs in the oldest quasars.  Note that
although the argument above relies on many simplifying assumptions,
all but one tend to overestimate the resulting SMBH mass,
so the SMBH masses in table \ref{tab_merger} should be
considered extreme upper limits.  Our assumptions are
summarized as follows.

1.  All halos contain central BHs at $0.02\%$ of the halo mass
(this may seriously overestimate the number of BHs).

2.  All relevant merging is completed by $z = 5$, instead of
today, so the number of infalling BHs is again exaggerated.

3.  In the $\Lambda$CDM world, going from redshift $20$ to $5$
allows over $1~{\rm{Gyr}}$ for merging, so the number of
more massive subhalos available to merge is artificially
large.  This overestimates the final SMBH mass, because
larger infalling subhalos will dynamically sink faster.

4.  Infall radii begin at $r_i \leq R_{\rm{s}}$, which for
some mergers represents a head start of multiple orders
of magnitude.

5.  The choice of an isothermal halo model probably
exaggerates the central density, artificially speeding
up the action of dynamical friction near the center.

6.  SMBH coalescence is assumed to occur arbitrarily quickly,
but in reality takes at least $\sim 10^8$ years, and probably
much longer \citep{milo01}.

7.  The choice to trace a $4 \times 10^{13}~\rm{M}_{\sun}$
halo today back in time tends to \emph{under}estimate the
number of  progenitor subhalos.    Extrapolating this mass linearly
from \citet{volo03}, we find that increasing it by $100$ times
would increase $dN/dM$ at $z = 5$ by a factor of $\lesssim 10$
while approximately maintaining its slope.

In this case, the largest halo at $z = 5$ would be
$\sim 10^{13}~\rm{M}_{\sun}$, so halo speeds should be
comparable to those in M87 today, the giant elliptical
at the center of the Virgo galaxy cluster.  Radial
velocities for globular clusters around M87 \citep{cohe00}
range from $1000$ to $2000~\rm{km~s^{-1}}$.  These changes,
using $v \approx 1000~\rm{km~s^{-1}}$, allow a SMBH to become
almost $10^9~\rm{M}_{\sun}$ in $0.7~\rm{Gyr}$ from an
initial radius $r_i = R_{\rm{s}} = 1~\rm{kpc}$, 
which is a factor of $\sim 4$ greater than our original
result.

So if we attempt to correct for this seventh
assumption, we approach (but do not reach) the
$3 \times 10^9~\rm{M}_{\sun}$ hole that quasar
observations require.  But this is only true if
we take literally the six highly conservative
assumptions above.

\subsection{Accretion} \label{accretion}

Consider first gas accretion, followed by feeding on dark matter.

\emph{Gas accretion} is probably the simplest and most
successful way to explain the growth of an IMBH into a SMBH, or
even a large stellar mass BH into a SMBH.  \citet{haim01} show that
if the black hole accretes at the Eddington limit continuously,
from birth through detection as a quasar, it achieves the needed
growth.  They give the Eddingtion luminosity,
\begin{equation}
L_{\rm{E}} = 4 \pi G M_{\bullet} c \mu_{\rm{e}} m_{\rm{p}}
\sigma_{\rm{T}}^{-1} \, ,
\label{Ledd}
\end{equation}
and introduce a radiative efficiency
$\epsilon \equiv L / \dot{M_{\bullet}} c^2$
and a fraction of Eddington output
$\eta \equiv L / L_{\rm{E}}$.
The resulting $e$-folding time is
\begin{equation}
t_{\rm{acc,e}} = \frac{M_{\bullet}}{\dot{M_{\bullet}}} = 
4 \times 10^7 \left( \frac{\epsilon}{0.1} \right)
\eta^{-1} \, \rm{yr} \, ,
\label{e-fold}
\end{equation}
which equals a $10$-folding timescale
\begin{equation}
t_{\rm{acc,10}} = 9.2 \times 10^7 \left( \frac{\epsilon}{0.1}
\right) \eta^{-1} \, \rm{yr} \, .
\label{10-fold}
\end{equation}
For a fiducial value of $\epsilon \approx 0.1$ and continuous
Eddington-limited accretion ($\eta \approx 1$), one can
calculate the initial BH mass needed to grow to
$3 \times 10^9~\rm{M}_{\sun}$ in the alloted time.

If the seed BH is primordial, it might accrete steadily from
approximately the beginning of time.  In this case, it has
$0.88~\rm{Gyr}$ to grow; that's $9.6$ orders of magnitude in
mass by eq. \ref{10-fold}.  So the initial seed mass must
have been $M_{\rm{seed}} \gtrsim 0.8~\rm{M}_{\sun}$,
which, interestingly, could have been formed during the
cosmological quark-hadron phase transition \citep{jeda99}.

If the seed BH is a normal population III stellar
remnant, or a supermassive star's remnant, then it had
$0.7~\rm{Gyr}$.  It's initial mass was then
$M_{\rm{seed}} \gtrsim 70~\rm{M}_{\sun}$, which is
completely plausible for the first generation of stars.
One star of $\gtrsim 260~\rm{M}_{\sun}$ would
suffice \citep{hege02}.

In $0.6~\rm{Gyr}$, the seed BH mass needs to be
$M_{\rm{seed}} \gtrsim 900~\rm{M}_{\sun}$.  This
applies for the case where population III stars must
sink to the center of their parent halos and merge
before beginning to grow significantly by gas
accretion.

In the case where the seed BH formed from the evolution
of normal population III stars, the mass of the
seed star is suggestively close to what one would expect.
Thus this route is possible, but only if Eddington limited
accretion is maintained during the entire IMBH growth
process.  If the seed hole evolved from a more massive SMS,
then there is time to spare, and the accretion rate could
have dropped for some of the growth period.

The success of this quick exponential growth calculation
seems to indicate that baryon accretion is the most realistic
mechanism for IMBH growth.  When the BH is small, the
Eddington mass accretion rate is not prohibitively large,
and by $z = 6.41$ we have very strong evidence that the
quasar is accreting at its Eddington rate \citep{will03}.
So it is reasonable to speculate that it has been
accreting at this rate all along.

However it is worth noting that the central SMBHs in modern
galaxies appear to grow by a factor of 10 on timescales
of $\gtrsim 8~\rm{Gyr}$ \citep{merr00}.
This is very slow compared to Eddington limited growth,
requiring $\eta \lesssim 0.01$.  This dramatic drop in
accretion rate needs explaining, especially if we are to
claim that no similar drop has ever happened before $z \sim 6$
in all of cosmic history, as the notion of SMBH growth
by gas accretion onto population III stellar corpses
would mandate.  This problem is only alleviated if the
first stars were supermassive, or the seed BHs were
primordial and significantly larger than $1~\rm{M}_{\sun}$.

\emph{Accreting dark matter} is another possibility.
\citet{macm02} have proposed an interesting way to account
for IMBH growth in which the hole grows proportionally to
the dark matter halo as matter falls in.
As mentioned earlier in section \ref{flowchart}, this
assumption leads to the observed galaxy power spectrum
and the observed $M_{\bullet} - \sigma$ relation.

In numerical simulations \citep{macm03}, dark
matter was found to form a self-similar region surrounding
the BH, which does grow in proportion to the BH growth at the
center.  The formation of this region proceeds by
gravitational interactions between clumps of dark
matter (``particles'' in the simulations)
which must have formed earlier in the CDM universe.
The BH grows whenever the particles cross inside the
Schwarzschild radius.  (The BH growth rate is thus
understated in the sense that $R_{\rm{sch}}$ was used
instead of the larger capture radius.)

In the absence of angular momentum, the black hole
grows rapidly:  $M_{\bullet} \propto t^4$.
But for realistic departures from pure spherically
symmetric infall, in which the dark matter has angular
momentum to help it resist falling into the hole,
the growth of the BH is only linear in time:
$M_{\bullet} \propto t$ (although the
$M_{\bullet} - \sigma$ relation is recovered
either way).

The simulations were not performed in standard units,
and therefore cannot be immediately applied to relevant
halo sizes and cosmic timescales.  Here we simply note
that the simulated BH growth is slow (e.g., linear in time,
as compared with exponential in time for gas accretion).
One therefore needs an unusually large IMBH to start
from, which requires an unusually large SMS as its origin.

\section{Conclusions}  \label{conclusions}

In this paper we have surveyed the state of knowledge of
SMBH formation and growth in light of a difficult new data
point to satisfy:  a quasar at $z = 6.41$ whose SMBH was
$3 \times 10^9~\rm{M}_{\sun}$ when the universe was only
$880~\rm{Myr}$ old.  We have extended the calculations of
others on SMBH growth via mergers and accretion.

In the case of SMBH growth by mergers, we follow the
output from a CDM simulation in order to track progenitor
subhalos.  We stipulate that a subhalo must sink to the
center by dynamical friction after a merger within the
time available for SMBH growth.  This limits the mass of
subhalos and therefore BHs which could have coalesced
in time, thus constraining the final SMBH mass.  We find
that in any realistic case, mergers are incapable of
growing a BH with the needed speed.

Accreting baryonic matter represents a viable growth process.
For this mechanism, we impose time limits sensitive to the
manner in which the seed BH first formed.  The following
possibilities emerge:

1.  A primordial black hole that formed during the quark-hadron
phase transition has accreted at the Eddington rate continuously
ever since baryons have been able to cluster until at least
the quasar epoch.

2.  A large population III seed star with
$M \approx 260~\rm{M}_{\sun}$ evolved to a BH,
and accreted at the maximum (Eddington) rate continuously
between its formation and the quasar epoch.

3.  A collection of merged population III stellar black
holes formed an IMBH with $M \approx 900~\rm{M}_{\sun}$, and
accreted at the Eddington rate continuously through the
quasar epoch.

Each of these options involves seed stars at
the upper end of their predicted mass range.
So although the Eddington limited gas accretion scenario
is adequate to grow the holes in the time available,
it is not entirely satisfying, because it requires
heavy gas infall which is never significantly diminished
before the quasar era, despite the relative gas
suffocation which has evidently happened since.
One can avoid imposing this requirement by starting
with a larger BH:

4.  A single SMS, or the merged remains of a dense cluster
of stars, or a PBH formed well after the quark-hadron phase
transition, yielded an IMBH much larger than
$\sim 100~\rm{M}_{\sun}$.  It accreted through the
quasar epoch, but did \emph{not} need to maintain the
maximum accretion rate during that time.

One other option, wherein the BH grows in proportion
to its host halo by eating dark matter,
is expected to yield slower BH growth.
As such, normal population III stellar seeds would be
inadequate, although a particularly massive SMS seed
remains a possibility.

We conclude by suggesting that \emph{some} unconventional
mechanism is needed to realistically beat the clock
for the early formation of SMBHs.  BH growth through mergers
is too slow, and growth by accreting nonstop at the
Eddington limit is probably too contrived.  The remedy
appears to require one of the following:  (a) PBHs
from unusually large overdensity pockets, (b)
population III starbursts well in excess of those
predicted by current simulations, or
(c) supermassive stars.

\acknowledgments

This work was supported by NASA under the
Colorado Space Grant Consortium.  We are grateful to
R. N. Henriksen and J. D. MacMillan for helpful
correspondence.

\end{document}